\documentclass[aps,prl,twocolumn,showpacs,preprintnumbers,amsmath,amssymb,superscriptaddress]{revtex4}%

\usepackage{graphicx}%
\usepackage{dcolumn}
\usepackage{amsmath}
\usepackage{color}
\usepackage{bm}

\makeatletter
\def\btt#1{\texttt{\@backslashchar#1}}%
\DeclareRobustCommand\bblash{\btt{\@backslashchar}}%
\makeatother

\begin{document}

\preprint{PREPRINT (\today)}

\title{Anomalous electron-phonon coupling probed on the surface of
ZrB$_{12}$ superconductor. }

\author{R.~Khasanov}
\affiliation{ Laboratory for Neutron Scattering, ETH Z\"urich and
Paul Scherrer Institut, CH-5232 Villigen PSI, Switzerland}
\affiliation{DPMC, Universit\'e de Gen\`eve, 24 Quai
Ernest-Ansermet, 1211 Gen\`eve 4, Switzerland}
\affiliation{Physik-Institut der Universit\"{a}t Z\"{u}rich,
Winterthurerstrasse 190, CH-8057, Z\"urich, Switzerland}
\author{D. Di~Castro}
\affiliation{Physik-Institut der Universit\"{a}t Z\"{u}rich,
Winterthurerstrasse 190, CH-8057, Z\"urich, Switzerland}
\affiliation{INFM-Coherentia and Dipartimento di Fisica,
Universita' di Roma "La Sapienza", P.le A. Moro 2, I-00185 Roma,
Italy} %
\author{M.~Belogolovskii}
\affiliation{Donetsk Physical and Technical Institute, National
Academy of Science of Ukraine, 83114 Donetsk, Ukraine}
\author{Yu.~Paderno}
\affiliation{Institute for Problems of Materials Science, National
Academy of Science of Ukraine, 03680 Kiev, Ukraine}
\author{V.~Filippov}
\affiliation{Institute for Problems of Materials Science, National
Academy of Science of Ukraine, 03680 Kiev, Ukraine}
\author{R.~Br\"utsch}
\affiliation{Laboratory for Material Behaviour, Paul Scherrer
Institut, CH-5232 Villigen PSI, Switzerland}
\author{H.~Keller}
\affiliation{Physik-Institut der Universit\"{a}t Z\"{u}rich,
Winterthurerstrasse 190, CH-8057, Z\"urich, Switzerland}

\begin{abstract}
Magnetization measurements under hydrostatic pressure up to
10.5~kbar in zirconium dodecaboride ZrB$_{12}$ superconductor
($T_c\simeq6.0$~K at $p=0$) were carried out. A negative pressure
effect on $T_c$ with ${\rm d} T_c/{\rm d} p =-0.0225(3)$~K/kbar
was observed. The electron-phonon coupling constant
$\lambda_{el-ph}$ decreases with increasing pressure with ${\rm
d}\ln\lambda_{el-ph}/{\rm d}p\simeq-0.20$\%/kbar. The magnetic
field penetration depth $\lambda$ was studied in the Meissner
state and, therefore, probes only the surface of the sample. The
absolute values of $\lambda$ and the superconducting energy gap at
ambient pressure and zero temperature were found to be
$\lambda(0)=$140(30)~nm and $\Delta_0=$1.251(9)~meV, respectively.
$\Delta_0$ scales linearly with $T_c$ as
$2\Delta_0/k_BT_c=4.79(1)$.  The studies of the pressure effect on
$\lambda$ reveal that $\lambda^{-2}$ increases with pressure with
${\rm d}\ln\lambda^{-2}(0)/{\rm d}p=0.60(23)$~\%/kbar. This effect
can not be explained within the framework of conventional
adiabatic electron-phonon pairing, suggesting that close to the
surface, an unconventional {\it non-adiabatic} character of the
electron-phonon coupling takes place.

\end{abstract}

\maketitle


The traditional concept of superconductivity is strictly
associated with the electron-phonon interaction. The conventional
theory is based on the Migdal-Eliashberg adiabatic approximation
\cite{Migdal58} that, in fact, leads to the prediction of many
peculiar features which are a direct evidence of a phonon mediated
superconductivity. The adiabatic approximation is valid if the
parameter $\omega_0/E_f $ is small ($\omega_0$ is the relevant
phonon frequency and $E_f$ is the Fermi energy). Usually this
parameter is regarded as a measure of nonadibaticity. However,
crossover from a conventional adiabatic to an unconventional
nonadiabatic regime does not depend only on the value of the
$\omega_0/E_f $ ratio. Paci {\it et al.} \cite{Paci05} show that
even in a case of small "adiabatic" ratio one would expect the
nonadiabatic coupling in superconductors having high value of the
electron-phonon coupling constant $\lambda_{el-ph}$. Among BCS
superconductors the zirconium dodecaboride (ZrB$_{12}$) is
probably a candidate for the observation of such type of anomalous
coupling. It stems from the rather small value of the Fermi energy
$\sim$1~eV \cite{Daghero04} that, together with the Debye
temperature $\sim$20~meV \cite{Lortz05}, leads to a ratio
$\omega_0/E_f\sim$0.02. A strong coupling ratio
$2\Delta/k_BT_c\simeq4.8$ was observed by surface sensitive
techniques \cite{Daghero04,Tsindlekht04}. This suggests that the
electron-phonon coupling constant, which has a bulk value
$\lambda_{el-ph}\simeq0.67$ \cite{Daghero04}, increases at the
surface. From the comparison with strong coupled metallic
superconductors \cite{Carbotte90} one would expect
$\lambda_{el-ph}^{surf.}\simeq1.7-1.9$. Moreover, it was pointed
out by Cappelluti {\it et al.} \cite{Cappelluti05} that
nonadiabatic character can be further enhanced by low charge
carrier density, that is the case for ZrB$_{12}$
\cite{Lortz05,Matthias68}.

One of the key feature of nonadiabatic superconductivity is the
observation of unconventional isotope  and pressure effects on the
magnetic field penetration depth $\lambda$. Note, that in
adiabatic superconductors (or in the superconductors where the
nonadiabatic effects are small) the pressure effect (PE)
\cite{Khasanov04,DiCastro05} as well as the isotope effect (IE)
\cite{DiCastro04} on $\lambda$ was found to be almost negligible
in comparison with substantial PE \cite{Khasanov05} and IE
\cite{Zhao97Hofer00Khasanov04aKhasanov04b} on $\lambda$ observed
in highly nonadiabatic high-$T_c$ cuprates. In this paper we
report on PE on $T_c$ and $\lambda$ studies in ZrB$_{12}$
superconductor. The magnetic penetration depth measured in the
Meissner state is largely determined by the surface
characteristics. The absolute value of $\lambda$ at zero
temperature and zero pressure was found to be
$\lambda(0)=140(30)$~nm. The transition temperature $T_c$ and the
electron-phonon coupling constant decrease with pressure with the
pressure effect coefficients $dT_c/dp=-0.0225(3)$~K/kbar and ${\rm
d}\ln\lambda_{el-ph}/dp\simeq-0.2$\%/kbar, respectively. In
contrast to $T_c$, $\lambda^{-2}(0)$ was found to increase with
${\rm d}\lambda^{-2}(0)/{\rm d}p=0.29(11)$~$\mu$m$^{-2}$/kbar.
Only a small part of this effect can be explained by a pressure
induced renormalization of the electron-phonon interaction and the
band structure changes. The major part is probably a consequence
of {\it nonadibatic} coupling of the charge carriers to the
crystal lattice appearing in ZrB$_{12}$ close to the surface.


Details on the sample preparation for ZrB$_{12}$ can be found
elsewhere \cite{Paderno02}. The single crystal has been grounded
in mortar and then sieved via 10~$\mu$m sieve in order to obtain
small grains needed for determination of $\lambda$ from
magnetization measurements. The grain size distribution was
determined by analyzing scanning electron microscope (SEM)
photographs. The hydrostatic pressure was generated in a
copper--beryllium piston cylinder clamp especially designed for
magnetization measurements under pressure \cite{Straessle02}. The
sample was mixed with Fluorient FC77 (pressure transmitting
medium) with a sample-to-liquid volume ratio of approximately
$1/6$. The pressure dependence of $T_c$ was taken from a separate
set of magnetization experiments where a small piece of indium
[$T_{c}(p=0)=3.4$~K] with known $T_{c}(p)$ dependence was added to
the sample and both $T_c$'s of indium and ZrB$_{12}$ were
recorded.
The field--cooled (FC) magnetization measurements were performed
with a SQUID magnetometer in a field of $ 0.5$~mT at temperatures
between $1.75$~K and $10$~K. The absence of weak links between
grains was confirmed by the linear magnetic field dependence of
the FC magnetization, measured at $0.25$~mT, $0.5$~mT, and
$1.0$~mT for the highest and the lowest pressures  at $T=1.75$~K.

\begin{figure}[htb]
\includegraphics[width=1.0\linewidth]{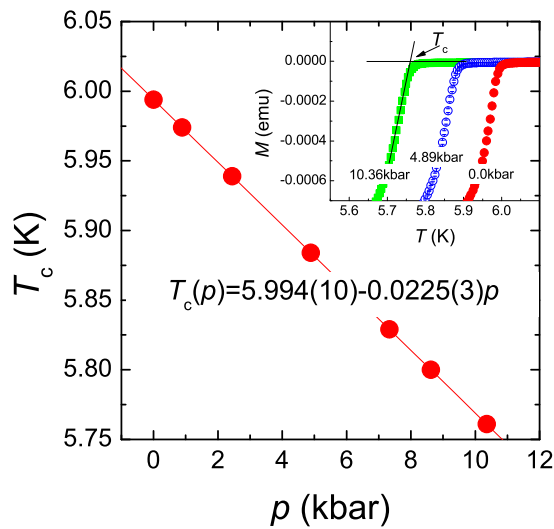}
\caption{Pressure dependence of the transition temperature $T_c$
for ZrB$_{12}$. The errors are smaller than the size of the
symbols. The inset shows $M(T)$ curves in the vicinity of $T_c$
for (from the left to the right) 10.36, 4.89, and 0.0~kbar.
 }
 \label{fig:Tc_vs_p}
\end{figure}

Figure~\ref{fig:Tc_vs_p} shows the pressure dependence of the
transition temperature $T_c$ of ZrB$_{12}$ obtained from
magnetization measurements. $T_c$ was taken from the linearly
extrapolated $M(T)$ curves in the vicinity of $T_c$ with $M=0$
line (see inset in Fig.~\ref{fig:Tc_vs_p}).
The linear fit yields $dT_c/dp=-0.0225(3)$~K/kbar. Note that this
value is in good agreement with $dT_c/dp\simeq-0.024$~K/kbar
obtained {\it indirectly} by Lortz {\it et al.} \cite{Lortz05}
from thermal expansion measurements.

The logarithmic volume derivative of $T_c$ in case of strong
coupled BCS superconductor can be described by the following
equation \cite{Tomita01}:
\begin{equation}
\frac{{\rm d}\ln T_c}{{\rm d}\ln V}=-B\frac{{\rm d}\ln T_c}{{\rm
d}p}= (2A-1)\gamma+A \frac{{\rm d} \ln \eta}{{\rm d}V}\ ,
 \label{eq:delta_Tc}
\end{equation}
where $A=1.04\lambda_{el-ph}[1+0.38\mu^\ast][\lambda_{el-ph} -
\mu^\ast(1+0.62\lambda_{el-ph})]^{-2}$ is a function of the
electron-phonon coupling constant $\lambda_{el-ph}$ and the
Coulomb pseudopotential $\mu^\ast$ \cite{Tomita01}, $B$ denotes
the bulk modulus, $\gamma=-{\rm d}\ln\langle\omega\rangle/{\rm
d}\ln V$ is the Gr\"uneisen parameter, $\langle\omega\rangle$ is
an average phonon frequency, $\eta\equiv N(E_f) \langle I^2
\rangle$ is the Hopfeld parameter \cite{Hopfeld71}. $N(E_f)$ is
the density of states at the  Fermi level, and $\langle I^2
\rangle$ is the average squared electronic matrix element. The
Hopfeld parameter $\eta$ generally increases under pressure with
${\rm d}\ln\eta/{\rm d}\ln V\approx-1$ for $s$-, and $p$-metal
superconductors \cite{Schilling92} and $-3$ to $-4$ for
transition-metal ($d$-electron) superconductors \cite{Hopfeld71}.
Assuming that ${\rm d}\ln\eta/{\rm d}\ln V\simeq-1$, $B=2490$~kbar
in analogy with UB$_{12}$ \cite{Dancausse92}, $\mu^\ast=0.1$ (that
is the typical value for conventional phonon-mediated
superconductors (see {\it e.g.} Ref.~\cite{Carbotte90}), and
taking $\lambda_{el-ph}\simeq0.67$ \cite{Daghero04}, for the
Gr\"uniesen parameter we get the value $\gamma\simeq2.83$. This
value is in reasonable agreement with $\gamma\simeq3.3$ obtained
at a temperature slightly above $T_c$ by Lortz {\it et al.}
\cite{Lortz05} based on thermal expansion measurements.

PE on the electron-phonon coupling constant $\lambda_{el-ph}$ can
be determined by using the well-known McMillan equation
\cite{McMillan68}
\begin{equation}
\lambda_{el-ph}\propto\frac{N(E_f)\langle I^2\rangle}{
\langle\omega^2\rangle},
 \label{eq:lambda_el-ph}
\end{equation}
which leads to
\begin{equation}
\frac{{\rm d}\ln \lambda_{el-ph}}{{\rm d}p}=-\frac{1}{B}\frac{{\rm
d}\ln\eta}{{\rm d}\ln V}-\frac{2\gamma}{B}.
 \label{eq:delta_lambda_el-ph}
\end{equation}
Substitution of $\gamma=2.83$ and ${\rm d}\ln\eta/{\rm d}\ln V=-1$
gives ${\rm d}\ln \lambda_{el-ph}/{\rm d}p=-0.19$\%/kbar. A
slightly larger value ${\rm d}\ln \lambda_{el-ph}/{\rm
d}p=-0.22$\%/kbar is obtained with $\gamma=3.3$ from the
Ref.~\cite{Lortz05}.

As a next step we studied the pressure effect on the magnetic
field penetration depth $\lambda$.
\begin{figure}[htb]
\includegraphics[width=1.0\linewidth]{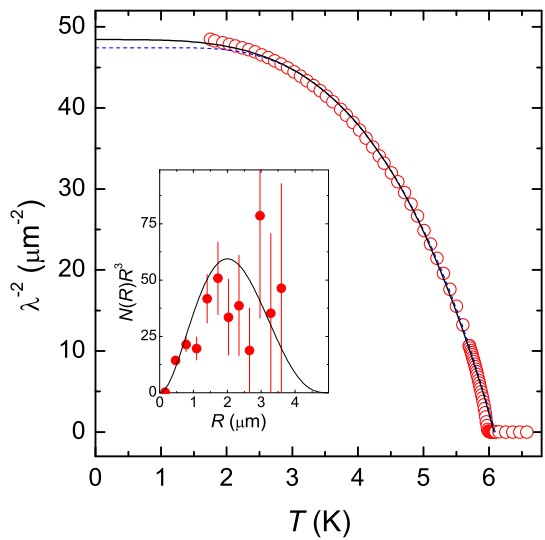}
\caption{ The temperature dependence of $\lambda^{-2}$ calculated
from the measured $\chi(T)$ by using
Eq.~(\ref{eq:Shoenberg-modified}). Lines represent fits with the
BCS model (dashed line), and with a power law (solid line). See
text for an explanation. The inset shows the volume fraction
distribution $N(R) R^3$ of the ZrB$_{12}$ powder determined from
SEM photographs. The errors are statistical. The solid line
represent the analytical $g(R)$ function used in
Eq.~(\ref{eq:Shoenberg-modified}). }
 \label{fig:lambda_vs_T}
\end{figure}
The temperature dependence of $\lambda$ was calculated from the
measured FC magnetization by using the Shoenberg formula
\cite{Shoenberg40}, modified for the known grain size distribution
$N(R)$ \cite{Porch93}:
\begin{eqnarray}
\chi & = & -\frac{3}{2}\int_0^\infty\left(1-
\frac{3\lambda}{R}\coth\frac{R}{\lambda}+
\frac{3\lambda^2}{R^2}\right)g(R){\rm d}R / \nonumber \\
& &\int_0^\infty g(R){\rm d}R \ \ \ ,
 \label{eq:Shoenberg-modified}
\end{eqnarray}
where $\chi=M/HV$ is the volume susceptibility, $V$ is the volume
of the sample, $R$ is the grain radius and $g(R)$ is the
analytical function describing the $N(R) R^3$ dependence (see
inset in Fig.~\ref{fig:lambda_vs_T}). The resulting temperature
dependence $\lambda^{-2}(T)$ at ambient pressure is shown in
Fig.~\ref{fig:lambda_vs_T}.
The reconstructed data were fitted with the empirical power-law
$\lambda^{-2}(T)/\lambda^{-2}(0)=1- (T/T_{c})^n$
\cite{Zimmermann95}. The fit yields
$\lambda^{-2}(0)=48.4(2)$~$\mu$m$^{-2}$, $T_c=6.078(5)$~K, and
$n=3.65(4)$.  Note, that the value of the power exponent $n$ is
close to ''4`` which corresponds to a strong--coupled BCS
superconductor \cite{Tinkham75}.

In order to
obtain the value of the superconducting gap $\Delta$, the
data have also been analyzed by means of the BCS model. For clean
superconductor the temperature dependence of $\lambda^{-2}$ can be
described in the following way \cite{Tinkham75}:
\begin{equation}
\frac{\lambda^{-2}(T)}{\lambda^{-2}(0)}=  1+
2\int_{\Delta(T)}^{\infty}\frac{\partial F}{\partial
E}\frac{E}{\sqrt{E^2-\Delta(T)^2}}\  dE \  ,
 \label{BCS-weak-coupled}
\end{equation}
where $F=(1+\exp(E/k_BT))^{-1}$ is  the Fermi function,
$\Delta(T)=\Delta_0\cdot \tilde{\Delta}(T/T_c)$ represents the
temperature dependence of the energy gap, and $\Delta_0$ is the
zero temperature value of the superconducting gap.
$\tilde{\Delta}(T/T_c)$ is the normalized gap taken from
Ref.~\cite{Muhlschlegel59}. The best fit to the data using
Eq.~(\ref{BCS-weak-coupled}) gives $T_c=6.09(2)$~K,
$\lambda^{-2}(0)=47.4(2)$~nm, and $\Delta_0=1.251(9)$~meV. The
ratio $2\Delta_0/k_BT_c=4.77(4)$ is found, suggesting that
ZrB$_{12}$ is a strong coupled BCS superconductor. Note, that a
rather close value $2\Delta_0/k_BT_c\simeq4.8$ has been obtained
in point-contact spectroscopy \cite{Daghero04} and tunnelling
\cite{Tsindlekht04} experiments. From the other hand a smaller
value $\simeq$3.7 has been reported by Lortz {\it et al.}
\cite{Lortz05} using the heat-capacitance technique, thus
suggesting a weak coupling strength. This difference has been
already pointed out by Tsindlekht {\it et al.}
\cite{Tsindlekht04}. It was explained by enhanced surface
characteristics of the ZrB$_{12}$ leading to rather different
superconducting properties of bulk
\cite{Lortz05,Tsindlekht04,Wang05} and surface
\cite{Daghero04,Tsindlekht04,Leviev05,Gasparov04}. Our
measurements were performed in the Meissner state, with the field
penetrating on a distance $\lambda$ from the surface and,
therefore, give a value of the superconducting gap consistent with
those one reported in the surface sensitive experiments
\cite{Tsindlekht04,Daghero04}.
To estimate the uncertainty in the {\it absolute} value of
$\lambda(0)$ we used a procedure similar to that one described in
Refs.~\cite{Khasanov04,Khasanov05}. The temperature dependence of
$\lambda(T)$ was calculated for $N(R)+\sqrt{N(R)}$ and
$N(R)-\sqrt{N(R)}$ distributions. The fit of the resulting
$\lambda(T)$ curves with the power law as well as with the BCS
model gives $\lambda(0)$ in the range from 110 to 170~nm.

\begin{figure}[htb]
\includegraphics[width=1.1\linewidth]{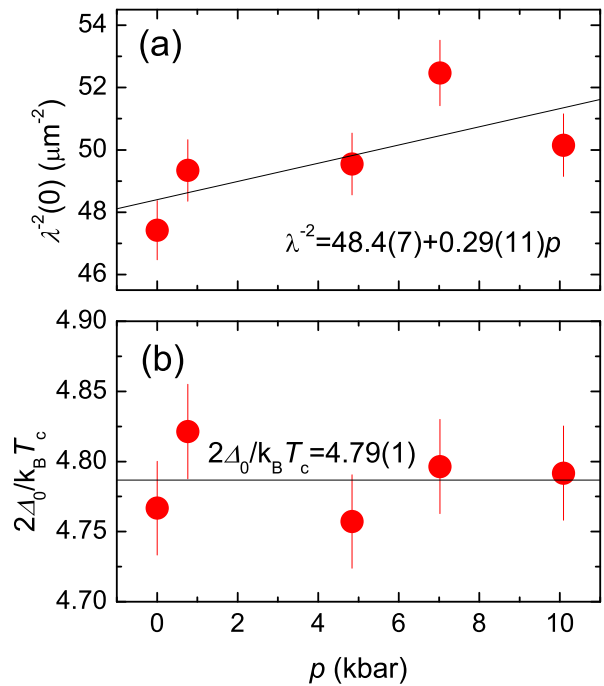}
\caption{ Pressure dependences of $\lambda^{-2}(0)$ (a) and
$2\Delta_0/k_BT_c$ (b). The solid lines are fits with parameters
shown in figures.}
 \label{fig:lambda-Delta_vs_p}
\end{figure}

Fig.~\ref{fig:lambda-Delta_vs_p}~(a) shows the pressure dependence
of $\lambda^{-2}(0)$ obtained by fitting the reconstructed
$\lambda(T)$ data at different pressures with the BCS model
[Eq.~(\ref{BCS-weak-coupled})]. In these experiments we studied
relative effects measured on the same sample in the same pressure
cell. The main systematic error of these measurements comes from
misalignments of the experimental setup occurring when the cell is
removed from the SQUID magnetometer, to change the pressure, and
put back again. This procedure was checked with a set of
measurements at constant pressure. The systematic scattering of
the magnetization data is about $0.3\%$, giving a relative error
in $\lambda^{-2}(T)$ of about 3\%. The reducing of the grain size
with pressure was taken into account in $\lambda(T)$ calculation
[Eq.~(\ref{eq:Shoenberg-modified})], by using the bulk modulus
reported above. The linear fit yields
$\lambda^{-2}(0)=48.4(7)+0.29(11)p$ implying that $\lambda^{-2}$
increases under pressure with ${\rm
d}\ln\lambda^{-2}(0)/dp=0.60(23)$\%/kbar [see
Fig.~\ref{fig:lambda-Delta_vs_p}~(a)].

To analyze the observed effect we used a procedure similar to that
one described by Di~Castro {\it et al.} \cite{DiCastro05}. There,
it was suggested that $\lambda^{-2}$ increases under pressure
because of two reasons: (i) band structure effects and (ii)
renormalization of the electron-phonon coupling \cite{DiCastro05}.
Under the assumption of ellipsoidal or cylindrical Fermi surface
the first one can be obtained as \cite{DiCastro05}
\begin{equation}
\frac{{\rm d}\ln\lambda^{-2}(0)}{{\rm d}p}=\frac{1}{3B}
-\frac{{\rm d}\ln N(E_f)}{{\rm d}p}\simeq \frac{1}{3B}
-\frac{1}{B}\frac{{\rm d}\ln \eta}{{\rm d}\ln V}.
 \label{eq:lambdaP_band}
\end{equation}
Here we used the fact that the pressure dependence of the
electronic matrix element $\langle I^2\rangle$ entering the
Hopfeld parameter $\eta$ can usually be neglected \cite{Lorenz04}.
Hence, by setting ${\rm d}\ln\eta/{\rm d}\ln V\simeq-1$, and
$B=2490$~kbar (see above) we obtain ${\rm
d}\ln\lambda^{-2}(0)/{\rm d}p=0.05$\%/kbar.

The electron-phonon renormalized penetration depth reduces to
$\lambda^{\ast-2}(0)=\lambda^{-2}(0)/(1+\lambda_{el-ph})$
\cite{Carbotte90}, where $\lambda(0)$ is the bare quantity we have
considered before. We have then:
\begin{equation}
\frac{{\rm d}\ln\lambda^{\ast-2}(0)}{{\rm
d}p}=-\frac{\lambda_{el-ph}}{1+\lambda_{el-ph}}\frac{{\rm d}\ln
\lambda_{el-ph}}{{\rm d}p}.
 \label{eq:lambdaP_BCS}
\end{equation}
By substituting $\lambda_{el-ph}\simeq0.67$ \cite{Daghero04} and
${\rm d}\ln \lambda_{el-ph}/{\rm d}p\simeq -0.2$\%/kbar obtained
above we get ${\rm d}\ln\lambda^{\ast-2}(0)/{\rm d}p\simeq
0.08$\%/kbar \cite{note1}. Thus the total pressure shift of
$\lambda^{-2}(0)$ expected assuming conventional (adiabatic)
coupling of the charge carriers to the lattice in ZrB$_{12}$ is of
the order of 0.13\%/kbar. This value is more than three times
smaller than the experimentally observed one 0.60(22)\%/kbar. This
implies that in addition to band structure effects and
renormalization of the electron-phonon coupling there are other
effects responsible for the increasing of $\lambda^{-2}(0)$ under
pressure. Bearing in mind that $\lambda$ measurements have been
performed in a Meissner state, the observed dependence of
$\lambda$ on $p$ can be explained assuming that in ZrB$_{12}$
close to the surface the coupling of the charge carriers to the
lattice has a {\it nonadiabatic} character. Note that similar
effect have been observed in YBa$_2$Cu$_4$O$_8$ that appears to be
a highly nonadiabtic superconductor \cite{Khasanov05}.

The results on the zero temperature superconducting gap $\Delta_0$
are summarized in Fig.~\ref{fig:lambda-Delta_vs_p}~(b), where the
ratio $2\Delta_0/k_BT_c$ is plotted as a function of the pressure
$p$. $\Delta_0$ and $T_c$ were obtained from the fit of
$\lambda^{-2}(T,p)$ data by using Eq.~(\ref{BCS-weak-coupled}).
The solid line represents a fit by the relation
$2\Delta_0/k_BT_c=const$ to the data. Bearing in mind that $T_c$
scales linearly with pressure (see Fig.~\ref{fig:Tc_vs_p}) the
constant ratio can be understood in the frame of the BCS theory,
which predicts $2\Delta_0/k_BT_c=3.52$. In the present study this
ratio was found to be pressure independent within experimental
errors, with mean value 4.79(1).

In conclusion, we performed magnetization measurements in
ZrB$_{12}$ under hydrostatic pressure. A {\it negative} pressure
effect on $T_c$ with ${\rm d} T_c/{\rm d} p =-0.0225(3)$~K/kbar is
observed. The electron-phonon coupling constant $\lambda_{el-ph}$
decreases with pressure with ${\rm d}\ln\lambda_{el-ph}/{\rm
d}p\simeq-0.20$\%/kbar. The magnetic field penetration depth
$\lambda$ measured in the Meissner state is largely determined by
the surface characteristics. $\lambda$ was found to increase with
pressure, with the pressure effect coefficient ${\rm
d}\ln\lambda^{-2}(0)/{\rm d}p=0.60(23)$\%/kbar. This coefficient
is much larger than that one estimated theoretically within the
adiabatic approximation. This can be explained by considering that
in ZrB$_{12}$, close to the surface, the coupling of the charge
carriers to the lattice has a {\it nonadiabatic} character. The
ratio $2\Delta_0/k_BT_c=4.79(1)$ is found to be pressure
independent and close to the strong coupling BCS value 4.8(1)
reported in Refs.~\cite{Daghero04,Tsindlekht04}. The value of
$\lambda$ extrapolated to zero temperature and at $p=0$ was
estimated to be 140(30)~nm.

The authors are grateful to S.~Str\"assle for help during
manuscript preparation. This work was supported by the Swiss
National Science Foundation and by the NCCR program
\textit{Materials with Novel Electronic Properties} (MaNEP)
sponsored by the Swiss National Science Foundation.


\begin{thebibliography}{99}
\bibitem{note1} Assuming for $\lambda_{el-ph}$ the surface
enhanced value 1.9 (see introduction) we obtain a slightly larger
value ${\rm d}\ln\lambda_{el-ph}/{\rm d}p\simeq0.13$\%/kbar.
\end{thebibliography}
\end{document}